# Folded Algebraic Geometric Codes From Galois Extensions

Ming-Deh Huang and Anand Kumar Narayanan *

October 30, 2018

## Abstract

We describe a new class of list decodable codes based on Galois extensions of function fields and present a list decoding algorithm. These codes are obtained as a result of folding the set of rational places of a function field using certain elements (automorphisms) from the Galois group of the extension. This work is an extension of Folded Reed Solomon codes to the setting of Algebraic Geometric codes. We describe two constructions based on this framework depending on if the order of the automorphism used to fold the code is large or small compared to the block length. When the automorphism is of large order, the codes have polynomially bounded list size in the worst case. This construction gives codes of rate $R$ over an alphabet of size independent of block length that can correct a fraction of $1 - R - \epsilon$ errors subject to the existence of asymptotically good towers of function fields with large automorphisms. The second construction addresses the case when the order of the element used to fold is small compared to the block length. In this case a heuristic analysis shows that for a random received word, the expected list size and the running time of the decoding algorithm are bounded by a polynomial in the block length. When applied to the Garcia-Stichtenoth tower, this yields codes of rate $R$ over an alphabet of size $(\frac{1}{\epsilon^2})^{O(\frac{1}{\epsilon})}$, that can correct a fraction of $1 - R - \epsilon$ errors.

*Computer Science Dept, University of Southern California (mdhuang,anand.narayanan@usc.edu)

# 1 Introduction

Error correction codes are combinatorial objects that are used in reliable transmission of information. In block error correction, a message which consists of $k$ symbols over an alphabet $\mathcal{S}$ is mapped into $N$ symbols over the alphabet. The image of this mapping that is contained in $\mathcal{S}^N$ defines a code. An element in the code is called a codeword and the Hamming distance between two codewords is defined as the number of coordinates where they differ. A received word is an arbitrary element in $\mathcal{S}^N$ that arises as a corrupted version of the image of a message. A decoder for the code tries to find the message transmitted from the corrupted received word. The integer $N$ is called as the block length of the code and $R = \frac{k}{N}$ the rate of the code.

A list decoder outputs the list of all codewords which have sufficient agreement with the received word. A list decodable code is said to correct $e$ errors if the number of codewords which are at a Hamming distance of at most $e$ from any received word is bounded by a polynomial in the block length of the code. There is a tradeoff between the rate and the fraction of errors ($\delta = \frac{e}{N}$) corrected for codes over an alphabet of size $q$ given by $R \leq 1 - H_q(\delta)$. Here $H_q(x) = x \log_q(\frac{q-1}{x}) + (1-x) \log_q(\frac{1}{1-x}))$ is the $q$-ary entropy function. Zyablov and Pinsker [19], proved the existence of list decodable codes whose parameters satisfy the above tradeoff with equality. In particular $\forall R, 0 < R < 1, \forall q \geq 2$ there exists list decodable codes of rate $R$ over an alphabet of size $q$ that can correct a fraction of $\delta = H_q^{-1}(1-R)$ errors. When the alphabet size $q$ is at least $2^{\frac{1}{\epsilon}}$, the fraction of errors corrected turns out to be at least $1 - R - \epsilon$. Observe that $R + \delta \leq 1$ is a fundamental bound. The list decodable codes of Zyablov and Pinsker approach this fundamental bound as the alphabet size gets larger. However the construction uses random coding arguments and the codes are not explicit. Guruswami and Rudra [10] described the first explicit family of codes called Folded Reed Solomon codes that achieve the $R + \delta \leq 1 - \epsilon$ trade off. We present an abstraction of their folding scheme to the setting of Galois extensions of function fields to give a new class of codes called Folded Algebraic Geometric codes.

Reed Solomon codes with unique decoding can correct a fraction of $1 - \frac{R}{2}$ errors. The Guruswami-Sudan List Decoding algorithm for Reed Solomon codes improved the bound to $\delta = 1 - \sqrt{R}$ [11]. In [14], Parvaresh and Vardy introduced a new class of codes (Parvaresh-Vardy Codes) that could correct a fraction of $1 - mR^{\frac{m}{m+1}}$ errors, for an integer $m \geq 2$. For certain rates, these can correct more errors than Reed Solomon codes running the Guruswami-Sudan list decoding algorithm. Building on [14], Guruswami and Rudra [10] constructed Folded Reed-Solomon codes of rate $R$ that could correct $1 - R - \epsilon$ fraction of errors. Let $N$ be the block length. The Folded Reed Solomon codes have an alphabet size requirement of $(\frac{N}{\epsilon^2})^{O(\frac{1}{\epsilon^2})}$, which is a large polynomial in the block length. Contained in [10] is a scheme to reduce the alphabet size based on concatenating Folded Reed Solomon codes with appropriate inner codes. Guruswami and Pathak [9] provide a generalization of the Parvaresh-Vardy code to the Algebraic-Geometric setting thereby reducing the alphabet size. By generalizing Folded Reed Solomon codes to Folded Algebraic Geometric codes we present a purely algebraic means of achieving the rate error correction tradeoff with alphabet size independent of the block length. Independent of this work, Guruswami [8] generalized Folded Reed-Solomon codes to codes from cyclotomic function fields that have an alphabet size that grows logarithmically in the block length.

Certain elements (automorphisms) from the Galois group of function field extensions are used to induce an ordering on the places of the function field used for defining the code. This ordering is used to fold the code and is exploited at the receiver to perform better error correction. Based on this general framework, we present two different construction depending on if the order of the automorphism used has order large or small compared to the block length. We present a list decoding algorithm for each case. The decoding algorithms are based on the interpolate and root find strategy common to [11][14][10][9]. The root finding step turns out to be much more complicated.

When the automorphism has an order comparable to the block length of the code, the list size is bounded by a polynomial in the block length. When applied to asymptotically optimal function fields towers that contain a large automorphism, the resulting codes of rate $R$ over an alphabet independent of the block length can correct a fraction of $1-R-\epsilon$ errors. However it is not known if such field extensions exists and we pose an open problem (See § 6) regarding such field extensions.

When the order of the automorphism used is small compared to the block length, the list decoding is much more complicated. We translate the root finding problem over the function field into a root finding problem over the local completion at a place where the automorphism acts as the Frobenius. The interpolated multivariate polynomial is mapped to one of a finite collection of polynomials in the local completion. We present an algorithm to solve the root finding problem over the local completion and a lifting of the solutions to the function field. The root finding algorithm in the local completion only depends on this finite collection of polynomials. If we pick a polynomial from this collection at random, the expected number of roots turns out to be polynomial in the degree of the interpolated polynomial and the size of the residue class field at that place. Under the heuristic that a random received word gets mapped to a random polynomial in this collection, the expected list size turns out to be bounded by a polynomial in the block length. (See § 4.2 for a discussion on why this heuristic assumption is reasonable.) When applied to the Garcia-Stichtenoth towers, we get codes over an alphabet of size $(\frac{1}{\epsilon^2})^{O(\frac{1}{\epsilon})}$ that can correct a fraction of $1-R-\epsilon$ errors. With our heuristic assumptions, the expected list size is bounded by $N^{O(\frac{1}{\epsilon})}$.

## 2 Folding Algebraic Geometric codes using elements from Galois Groups

In this section, we develop the ideas behind the code constructions and present a formal description of Folded Algebraic Geometric codes.

We begin by defining Reed-Solomon codes and then introduce Algebraic Geometric codes as generalizations of Reed-Solomon codes. Let $\mathbb{F}_q$ be the finite field with $q$ elements. Fix a size $N$ subset of the elements of the finite field $\mathbb{F}_q$. Messages are associated with polynomials $\{f \in \mathbb{F}_q[x], deg(f) < k\}$ with $k \leq N$. Here $deg(f)$ is the degree of the polynomial $f$. The image of $\{f \in \mathbb{F}_q[x], deg(f) < k\}$ under evaluation at this subset is the Reed-Solomon code. Observe that the alphabet size $q$ is at least as big as the block length for Reed-Solomon codes. Generalization to Algebraic Geometric codes yields codes of arbitrarily large block length over a fixed alphabet. Places in the function field take up the role of places of evaluation and the Riemann-Roch space takes up the role of the message space. We begin by stating some basic concepts in function fields. The reader is referred to [16] for a detailed description.

Let $K$ denote a function field that is a finite separable extension of the rational function field $\mathbb{F}_q(x)$, where $x$ is an indeterminate. Let $L/K$ be a finite Galois extension of $K$. It is assumed that both $L$ and $K$ have $\mathbb{F}_q$ as the field of constants. A ring $\mathcal{O} \subset L$ is called a valuation ring of the function field $L$ if $\mathbb{F}_q \subset \mathcal{O} \subset L$ and for all $f \in L$, either $f \in \mathcal{O}$ or $f^{-1} \in \mathcal{O}$. A valuation ring is a local ring and contains a unique maximal ideal. A place $v$ of the function field $L$ is defined as the maximal ideal of a valuation ring of $L$. If $v$ is a place, then the corresponding valuation ring is determined as $\mathcal{O}_v := \{f \in L : f^{-1} \notin v\}$. The quotient field $\mathbb{F}_v := \mathcal{O}_v/v$ is called the residue class field at $v$. The degree of the place $v$, denoted by $deg(v)$ is defined as the degree of the extension $\mathbb{F}_v$ over $\mathbb{F}_q$, and $v$ is called a rational place if the degree of $v$ is one. The natural reduction map $\mathcal{O}_v \longrightarrow \mathcal{O}_v/v$ is called as evaluation at $v$. Throughout, $f(v)$ denotes the evaluation of $f \in \mathcal{O}_v$ at $v$.

Let $\mathcal{V}_v(f)$ denote the valuation of $f$ at $v$ defined as follows. Let $t \in \mathcal{O}_v$ generate the ideal $u = <t>$. Any $f \in L$ can be written as $f = t^b f', b \in \mathbb{Z}$, where $f'$ is a unit in $\mathcal{O}_v$. The integer $b$ is independent of the choice of $t$ and is defined as $\mathcal{V}_v(f)$[16][I.1.11]. Let $S$ denote the set of places in $L$. The group of divisors is the additive free abelian group $\mathcal{D}$ generated by the places of $L$. The elements of $\mathcal{D}$ are called as divisors. In particular, a divisor $D$ is of the form $D = \sum_{v \in S} n_v v$, where $n_v \in \mathbb{Z}$ and $n_v = 0$ for all but a finite set. The degree of the divisor is $deg(D) = \sum_{v \in S} n_v deg(v)$. A divisor of a function $f \in L$ is defined as $div(f) := \sum_{v \in S} \mathcal{V}_v(f) v$. Let

$$\mathcal{L}(D) = \{f \in L : div(f) + D \geq 0\} \bigcup \{0\}$$

denote the Riemann-Roch space associated with the divisor $D$. The dimension of the Riemann-Roch space is lower bounded as $dim(\mathcal{L}(D)) \geq deg(D) - g + 1$. Here $g$ is the genus of the function field. Further, if $deg(D) \geq 2g - 1$, then $dim(\mathcal{L}(H)) = deg(D) - g + 1$.

Let $S_r$ denote the set of rational places of $L$. Let $S_D \subseteq S_r$ be a subset of the rational places disjoint from $P_\infty$, where $P_\infty \in S$ is a point at infinity. Let $D$ and $H$ denote divisors defined as $H = (\alpha - 1)P_\infty$ and $D = \sum_{v \in S_D} v$. Here $\alpha$ is a positive integer. Without loss of generality, assume that the degree of $P_\infty$ is 1. Algebraic Geometric codes were introduced by Goppa [5] and are defined as follows. The messages are associated with functions in $\mathcal{L}((\alpha - 1)P_\infty)$ and the code is the image of the evaluation of $\mathcal{L}((\alpha - 1)P_\infty)$ at the places of $S_D$ (Refer to [5] and [16] for a detailed description).

The minimum distance $d_{min}$ of Algebraic Geometric Codes is lower bounded by $d_{min} \geq \#S_D - deg(H)$ and likewise the dimension of the code (call $k$) by $k \geq deg(H) - g + 1$. The block length $\#S_D$ is upper bounded by the number of rational points in $L$. The number of rational points $N_L$ of a function field $L$ satisfies $\frac{N_L}{g} \leq \sqrt{q} - 1$(Drinfeld-Vladut Bound). If $q$ is a perfect square, then there exists function fields for which the number of rational points attains the upper bound [17]. An explicit construction of such function fields is presented in [4]. One can thus construct Algebraic-Geometric codes on these function fields of arbitrarily large block length over a constant alphabet $q$ such that both the rate and the relative minimum distance ($\frac{d_{min}}{N_L}$) are bounded away from zero.

## 2.1 Folded Algebraic Geometric Codes

In Folded Reed-Solomon codes that achieve list decoding capacity [10], the ordering of places was exploited by the decoder to get far better error correction. However, it was not apparent as to whether these techniques generalized to the case of Algebraic Geometric codes. We present such a folding scheme for Algebraic Geometric codes defined over certain Galois extensions.

Consider Reed-Solomon codes where all the elements of the multiplicative group of $\mathbb{F}_q$ are used for evaluation. The multiplicative group of a finite field is cyclic. Let $\gamma \in \mathbb{F}_q^*$ be a generator. In Folded Reed-Solomon codes the places of evaluation are enumerated as $1, \gamma, \gamma^2, \ldots, \gamma^{q-1}$. The evaluation of a polynomial $f$ at $\gamma^i$, gives us some information about the evaluation of $f$ at $\gamma^{i+1}$. This is exploited at the decoder [10]. We use the action of an element of the Galois group to induce an ordering of the places. First, we build some notation regarding Galois groups.

Let $Gal(L/K)$ denote the Galois group of the extension. The cardinality of $Gal(L/K)$ is $[L : K]$, where $[L : K]$ denotes the degree of the extension. For a place $v \in S$ and $\sigma \in Gal(L/K)$, let $\sigma(v) = \{\sigma(f) : f \in v\}$. Then $\sigma(v)$ is also a place in $L$ [16][Lem III 5.2]. Thus $Gal(L/K)$ acts on the places of $L$. This action can be naturally extended to divisors, so that the action of $\sigma \in Gal(L/K)$ on a divisor $D = \sum_{v \in S} a_v v$ is defined by $\sigma(D) = \sum_{v \in S} a_v \sigma(v)$. An element $\sigma \in Gal(L/K)$ induces an isomorphism on the residue fields of $v$ and $\sigma(v)$, given by $\sigma(f(v)) := \sigma(f)(\sigma(v))$. Thus $deg(v) = deg(\sigma(v))$. If $\sigma$ fixes the divisors $D$ and $H$, that is $\sigma(D) = D$ and $\sigma(H) = H$, then $\sigma$ defines an automorphism on the Algebraic Geometric code [16][VIII.3].

Let $v$ and $v'$ denote two places in $L$ such that $\sigma^{-1}(v) = v'$. Let $f \in L$ be an arbitrary function.

$$\sigma(f(v')) = \sigma(f)\sigma(v')$$
$$= \sigma(f)\sigma(\sigma^{-1}(v))$$
$$= \sigma(f)(v)$$

Thus from the evaluation of $f$ at $v'$ we can infer the evaluation of $\sigma(f)$ at $v$. We now order the places of evaluation of the code so that this can be exploited at the decoder.

For a place $v \in L$, an automorphism $\sigma \in Gal(L/K)$ and a positive integer $m'$, define $\Gamma_\sigma^{m'}(v)$ to be the ordered set $\{v, \sigma^{-1}(v), \ldots, \sigma^{-m'+1}(v)\}$. The evaluation of a function $f \in L$ at $\Gamma_\sigma^{m'}(v)$ is defined as $f(\Gamma_\sigma^{m'}(v)) := \{f(v), f(\sigma^{-1}(v)), \ldots, f(\sigma^{-m'+1}(v))\}$. Observe that $f(\Gamma_\sigma^{m'}(v)) \in \bigoplus_{i=0}^{m'-1} \mathbb{F}_{\sigma^{-i}(v)}$.

Let $b = [L : K]$ denote the degree of extension and $\sigma \in Gal(L/K)$ be of order $m$ in $Gal(L/K)$. Let $u$ be a place in $K$ that splits completely in the extension $L/K$. Then for every place $v$ above $u$, $\sigma^i(v)$ are all distinct for $i = 0, \ldots, m-1$. Thus $\Gamma_\sigma^m(v)$ consists of distinct places. Hence the set of places lying above $u$ in $L$ is partitioned into $\frac{b}{m}$ cycles under the action of $\sigma$ (and $\sigma^{-1}$ as well) with each cycle of length $m$. Such an element in the Galois group of order $m$ will be used to get a Folded Algebraic Geometric code with the folding parameter $m$. The set of places used to define the code is restricted to the set of rational places that resulted out of complete splitting in the extension.

## 2.2 Code Definition, Encoding and Parameters

We now formally describe the encoding process. Let $S_{sp}$ denote the set of rational places in $L$ that resulted out of complete splitting and with support disjoint from points at infinity. Denote the cardinality of $S_{sp}$ by $n$. Observe that as $v$ resulted out of splitting, $\Gamma_\sigma^m(v)$ represents a cycle of distinct places under the action of $\sigma^{-1}$. Then $S_{sp}$ is partitioned into $N := \frac{n}{m}$ cycles under the action of $\sigma^{-1}$. In particular $S_{sp} = \{\Gamma_\sigma^m(v_1), \Gamma_\sigma^m(v_2), \ldots, \Gamma_\sigma^m(v_N)\}$. Here $S_{rep} := \{v_1, v_2, \ldots, v_N\}$ is a fixed set of representatives of the orbits (cycles) of places in $S_{sp}$ under the action of $\sigma^{-1}$.

In the Folded AG code, $N = \frac{n}{m}$ will be the block length of the code. Let $H = (\alpha - 1)P_\infty$ be a divisor in $L$, where $P_\infty$ is a rational point at infinity in $L$ fixed by $\sigma$. Let $\mathcal{L}((\alpha - 1)P_\infty)$ denote the Riemann-Roch space associated with the divisor $G$. Here $\mathcal{L}((\alpha - 1)P_\infty)$ constitutes the message space and any function $f \in \mathcal{L}((\alpha - 1)P_\infty)$ is encoded as follows. The codeword corresponding to message $f \in \mathcal{L}((\alpha - 1)P_\infty)$ is the evaluation of $f$ at $S_{sp}$. The folded code is viewed as a code over an alphabet $q^m$.

In particular, the codeword is $\{f(\Gamma_\sigma^m(v_1)), f(\Gamma_\sigma^m(v_2)), \ldots, f(\Gamma_\sigma^m(v_N))\}$. The rate of the code depends on the dimension $k := dim(\mathcal{L}((\alpha - 1)P_\infty))$. The rate of the code $R = \frac{k}{mN} = \frac{k}{n}$.

## 2.3 A generalization with arbitrary folding

We now describe a variant of the Folded Algebraic Geometric Codes suited to case where the automorphism $\sigma$ used for folding has a large order. Let $m' < m$ be a positive integer. Without loss of generality assume that $m'$ divides $m$. Recall that $m$ is the order of $\sigma$ in $Gal(L/K)$. The cycle $\Gamma_\sigma^m(v)$ can be further partitioned into $\frac{m}{m'}$ ordered sets $\{\Gamma_\sigma^{m'}(v), \Gamma_\sigma^{m'}(\sigma^{-m'}(v)), \ldots, \Gamma_\sigma^{m'}(\sigma^{m-m'}(v))\}$. This way we can partition $S_{sp}$ into $\frac{n}{m'}$ disjoint ordered set of places each of size $m'$.

$S_{sp} = \{\Gamma_\sigma^{m'}(v_1), \Gamma_\sigma^{m'}(\sigma^{-m'}(v_1)), \ldots, \Gamma_\sigma^{m'}(\sigma^{m-m'}(v_1)), \Gamma_\sigma^{m'}(v_2), \Gamma_\sigma^{m'}(\sigma^{-m'}(v_2)), \ldots, \Gamma_\sigma^{m'}(\sigma^{m-m'}(v_2))$
$, \ldots, \Gamma_\sigma^{m'}(v_{\frac{n}{m}}), \Gamma_\sigma^{m'}(\sigma^{-m'}(v_{\frac{n}{m}})), \ldots, \Gamma_\sigma^{m'}(\sigma^{m-m'}(v_{\frac{n}{m}}))\}.$

By evaluating $f \in \mathcal{L}((\alpha - 1)P_\infty)$ at $S_{sp}$ partitioned in this way gives us Folded codes over an alphabet of size $q^{m'}$ of block length $N' := \frac{n}{m'}$ and rate $\frac{k}{n}$. Observe that the first construction is a special case of the second construction with $m = m'$.

# 3 List Decoding Folded Algebraic Geometric Codes

We describe a list decoding algorithm for the Folded Algebraic Geometric codes in this section. The decoding algorithm proceeds by first interpolating a multivariate polynomial based on the received word. The basic outline of the algorithm is similar to the second decoding algorithm presented in [9], though the steps in the algorithm are considerably more complicated.

## 3.1 Building the Multivariate Interpolation Polynomial

We describe the interpolation algorithm for the code construction where $m' = m$ and later describe the generalization. The multivariate interpolation step is essentially identical to [9]. Let $\{Y_j, v_j \in S_{rep}\}$ denote the received word. Here $Y_j \in \bigoplus_{i=0}^{m} \mathbb{F}_{\sigma^{-i}(v_j)}$. Let $\{y_v, v \in S_{sp}\}$ where $y_v \in \mathbb{F}_q$ denote the corresponding unfolded received word. Find a non zero multivariate polynomial $Q \in L[z_1, z_2, \ldots, z_m]$ such that

- $\forall f_1, f_2, \ldots, f_m \in \mathcal{L}((\alpha-1)P_\infty)$, we require $Q(f_1, f_2, \ldots, f_m) \in \mathcal{L}(lP_\infty)$

- $\forall v \in S_{sp}, \forall f_1, f_2, \ldots, f_m \in \mathcal{L}((\alpha-1)P_\infty)$ such that $f_1(v) = y_v, f_2(v) = y_{\sigma^{-1}(v)}, \ldots, f_m(v) = y_{\sigma^{-m+1}(v)}$, we require $\mathcal{V}_v(Q(f_1, f_2, \ldots, f_m)) \geq r$ where $l$ and $r$ are integer parameters determined later. Here, $r$ is the multiplicity parameter and $\mathcal{V}_v$ denotes the valuation at $v$.

The symbol corresponding to a place $v$ is said to be in agreement if the received symbol at $v$, $(y_v, y_{\sigma^{-1}(v)}, \ldots, y_{\sigma^{-m+1}(v)})$, is the actual transmitted symbol. The agreement parameter $T$ is defined as the number of locations (places in $S_{rep}$) at which there is an agreement. By construction we see that if the symbol corresponding to $v$ is in agreement, then so are the symbols corresponding to $\sigma^{-a}(v), 0 \leq a \leq m-1$. Thus for every symbol corresponding to a place $v \in S_{rep}$ that is in agreement, we get $m$ symbols corresponding to places $\sigma^a(v) \in S_{sp}, 0 \leq a \leq m-1$ that are in agreement. Define $t = Tm$.

**Lemma 3.1.** Let $rt > l$. If $f \in \mathcal{L}((\alpha-1)P_\infty)$ satisfies $f(v) = y_v, f(\sigma^{-1}(v)) = y_{\sigma^{-1}(v)}, \ldots, f(\sigma^{-m+1}(v)) = y_{\sigma^{-m+1}(v)}$ for at least $T$ of the places $v \in S_{rep}$, then $Q(f, \sigma(f), \ldots, \sigma^{m-1}(f)) = 0$.

*Proof* : Let $S_T \subseteq S_{rep}$ denote the set of places in $S_{rep}$ such that $f(v) = y_v, f(\sigma^{-1}(v)) = y_{\sigma^{-1}(v)}, \ldots, f(\sigma^{-m+1}(v)) = y_{\sigma^{-m+1}(v)}, \forall v \in S_T$. Observe that if $f(\sigma^{-i+1}(v)) = y_{\sigma^{-i+1}(v)}$, for $i = 1, \ldots, m$ for some $v \in S_{rep}$, then $f(v) = y_v, (\sigma(f))(v) = y_{\sigma^{-1}(v)}, \ldots, (\sigma^{m-1}(f))(v) = y_{\sigma^{-m+1}(v)}$. The cardinality of $S_T \geq T$, so $\sum_{v \in S_{sp}} \mathcal{V}_v(Q(f, \sigma(f), \ldots, \sigma^{m-1}(f))) \geq rmT = rt > l$. But $Q(f, \sigma(f), \ldots, \sigma^{m-1}(f)) \in \mathcal{L}(lP_\infty)$. This is because $\sigma$ fixes $P_\infty$ and thus $\sigma^j(f) \in \mathcal{L}((\alpha-1)P_\infty) \forall f \in \mathcal{L}((\alpha-1)P_\infty)$ and $j \in \mathbb{Z}$. Thus $Q(f, \sigma(f), \ldots, \sigma^{m-1}(f) = 0)$. □

In other words, any function (message) $f \in \mathcal{L}((\alpha-1)P_\infty)$ whose evaluation (codeword) has an agreement of at least $T$ with the received word satisfies $Q(f, \sigma(f), \ldots, \sigma^{m-1}(f)) = 0$.

The reader is referred to the original paper [9] for details regarding the construction of $Q$ and a discussion relating to representation needed to efficiently compute $Q$. The construction presented there runs in time polynomial in the block length. A multivariate polynomial $Q$ with the desired properties exists and can be constructed in polynomial time for the agreement parameter $T \geq \sqrt[m+1]{N(\alpha-1)^m}$ [9]. The multiplicity parameter satisfies $r := \left\lceil \frac{\alpha+g+m\sqrt[m+1]{N(\alpha-1)^m}}{t - \sqrt[m+1]{N(\alpha-1)^m}} \right\rceil$. We then set $l := rt - 1$. Moreover the degree $d$ of the multivariate polynomial $Q$ is upper bounded by $d \leq \frac{l-g}{\alpha-1}$ which at worst grows linearly in the block length.

For the general case of $m' \neq m$ we make some modifications to the interpolation algorithm. Let $s' < m'$ be an integer and for each $v \in S_{rep}$ which is in agreement, at least $m' - s' + 1$, $s'$-tuples satisfy $f(v') = y'_v, f(\sigma^{-1}(v')) = y_{\sigma^{-1}(v')}, \ldots, f(\sigma^{-m'+1}(v')) = y_{\sigma^{-m+1}(v')}$. These are $v' \in \{v, \sigma-1(v), \ldots, \sigma^{-s+1}(v)\}$. From an analysis analogous to [10] with the interpolation algorithm from [9], it is clear that an interpolation step can performed to obtain an $s'$ variate polynomial $Q$ that satisfies $Q(f, \sigma(f), \ldots, \sigma^{s'-1}(f)) = 0$ over $L$ for all codewords $f \in \mathcal{L}(\alpha - 1)P_\infty$ that have an agreement of $(\frac{m'}{m'-s'+1}(\frac{\alpha}{N})^{\frac{s'}{s'+1}})$ with the received word. Thus we can correct $N' - N'(\frac{m'}{m'-s'+1}(\frac{\alpha}{N})^{\frac{s'}{s'+1}}) = N' - N'(\frac{m'}{m'-s'+1}(R + \frac{m'g}{N})^{\frac{s'}{s'+1}})$ errors. The degree of the polynomial $Q$ is again bounded by $\frac{l-g}{\alpha-1}$.

## 3.2 Frobenius Elements and Ramification Groups

Here we describe certain concepts in Galois extensions on which the decoding algorithms depend. Let $v$ be an arbitrary place in $L$ that is above a place $u$ in $K$. The decomposition group of $v$ is defined as $\mathcal{D}_v := \{\sigma \in Gal(L/K) : \sigma(v) = v\}$. Thus the decomposition group of a place is the set of all elements in the Galois group that fix that place. For $\sigma \in \mathcal{D}_v$, the action of $\sigma$ on the residue class field $\mathbb{F}_v$ is well defined. That is $\sigma(f(v)) = \sigma(f)\sigma(v) = \sigma(f)(v)$. Thus, there is a natural homomorphism $\phi : \mathcal{D}_v \longrightarrow Gal(\mathbb{F}_v/\mathbb{F}_u)$. The homomorphism is surjective. The kernel of this homomorphism $\mathcal{I}_v$ is called as the inertia group of $v$. The following definition for the inertia group is equivalent $\mathcal{I}_v = \{\sigma \in Gal(L/K) : \sigma(f)(v) = f(v), \forall f \in \mathcal{O}_L\}$, where $\mathcal{O}_L$ denotes the ring of integers of $L$. When the place $v$ is totally ramified, the inertia group $\mathcal{I}_v$ is the whole Galois group $Gal(L/K)$. When the place $v$ is unramified, the inertia group is trivial. The residue class field extension $\mathbb{F}_v/\mathbb{F}_u$ is cyclic and is hence generated by a single element. Moreover if $v$ is unramified then $\mathcal{I}_v$ is trivial and hence there is a unique element $\sigma_v \in Gal(L/K)$, called the Frobenius element at $v$, such that $\sigma_v(f) = f^{\#(\mathcal{O}_u/u)} \mod v$ for all $f \in \mathcal{O}_L$.

Let $w$ be a place in $L$ above a place $u$ in $K$. The set of decomposition groups of places above $u$ are conjugates [15][Proposition 9.7]. Thus $Gal(L/K)/\mathcal{D}_w$ is the set of decomposition groups of places above $u$. Each of these decomposition groups are generated by the respective Frobenius elements of places above $u$. Denote by $\mathbb{H}_u := \{\sigma_w, w \text{ is a place above } u\}$. This set of all Frobenius elements of places in $L$ lying above $u$ is called as the Artin conjugacy class of $u$. Let $\Psi \subseteq Gal(L/K)$ be the conjugacy class of an arbitrary element in $Gal(L/K)$.

Tchebotarev Density Theorem ([13],[15][Thm 9.13B]) states that,

$$\left|\#\{u \in K : \mathbb{H}_u = \Psi\} - \frac{\#\Psi}{\#Gal(L/K)}\frac{q^{deg(u)}}{deg(u)}\right| \leq 2g(K)\frac{\#\Psi}{\#Gal(L/K)}q^{\frac{deg(u)}{2}} + \sum_{u \in L,\ e(u)>1} deg(u)$$

Here $g(K)$ denotes the genus of $K$ and $e(u)$ denotes the ramification index of $u$. From the Riemann-Hurwitz genus formula [16], it is evident that $\sum_{u \in L,\ e(u)>1} deg(u)$ grows at worst linearly in $g$ and $[L:K]$.

## 3.3 The Root Finding Problem

From the previous section, for the case of $m' = m$ it is evident that, messages that have an agreement of at least $T$ with the received word are a subset of $f \in \mathcal{L}((\alpha - 1)P_\infty)$ that satisfy $Q(f, \sigma(f), \sigma^2(f), \ldots, \sigma^{m-1}(f)) = 0$. Thus we can find all the messages in the list if we could enumerate all $f \in \mathcal{L}((\alpha - 1)P_\infty)$ that satisfy $Q(f, \sigma(f), \sigma^2(f), \ldots, \sigma^{m-1}(f)) = 0$. We have to solve the following root finding problem.

*Given a polynomial $Q \in L(z_1, z_2, \ldots, z_m)$ such that for every $h_1, h_2, \ldots, h_m \in \mathcal{L}((\alpha - 1)P_\infty)$, $Q(h_1, h_2, \ldots, h_m) \in \mathcal{L}(lP_\infty)$ and an automorphism $\sigma \in Gal(L/K)$, enumerate $f \in \mathcal{L}((\alpha - 1)P_\infty)$ that satisfy $Q(f, \sigma(f), \sigma^2(f), \ldots, \sigma^{m-1}(f)) = 0$.*

In the case where $m' \neq m$, the problem is to enumerate $f \in \mathcal{L}((\alpha - 1)P_\infty)$ that satisfy $Q(f, \sigma(f), \sigma^2(f), \ldots, \sigma^{s'-1}(f)) = 0$. We describe an algorithm that finds all such $f$. We handle the cases of the two code constructions separately. First we develop some notation common to both.

Let $w$ (unramified) be a place in $L$ lying above $u$ in $K$ such that $\sigma$ is the Frobenius element at $w$. Further assume that the degree of $u$ is $\eta = C \log_q(n)$, where $C$ is a positive constant. As $\sigma$ has order $m$ in $Gal(L/K)$, the degree of $w$ is $m\eta$. We recall that the action of $\sigma$ at $w$ is given by $\sigma(f) = f^{\#(\mathcal{O}_u/u)} \mod w$. That is $\sigma(f) = f^{q^\eta} \mod w$.

We now establish the existence of a place $w$ of degree $m\eta$ such that $\sigma$ is the Frobenius element at $w$. The existence follows from the Tchebotarev Density Theorem for function fields which gives the following lower bound on the number of $w$ of degree $m\eta$ such that $\sigma$ is the Frobenius at $w$, $\#\{w \in L : \sigma_w = \sigma, deg(w) = m\eta\} \geq \frac{1}{m}\frac{q^\eta}{\eta} \pm O(gq^{\frac{\eta}{2}})$. Thus for large enough $\eta$, such a place $w$ always exists. In fact, $\frac{1}{m}$ fraction of all unramified places of degree $\eta$ in $K$ have a place $w$ above it such that $\sigma$ is the Frobenius element at $w$. We are only interested in function fields where $n > g$. In this case, the choice of $\eta = C \log_q(n)$ with $C$ a large enough absolute constant, guarantees the existence of such a $w$. Moreover, such a place can be found in time polynomial in $n$ as follows. Exhaustively search through each place of degree $\eta$ in $K$, if there exists a place above it where $\sigma$ acts as the Frobenius.

## 4 Root Finding Step of the Decoding Algorithm

### 4.1 The easy case : $m' \neq m$ and $m$ large

The root finding problem is solved for the case where the automorphism used to fold has an order $m$, that is a constant fraction of $\frac{N'}{\log(N')}$. Further we assume that $m' < m$ is small and independent of the blocklength.

Let $w \in L$ be place of degree $m\eta$, where $\eta$ is the smallest integer such that $m\eta \geq \alpha$.

**Lemma 4.1.** *The evaluation map $\mathcal{L}((\alpha - 1)P_\infty) \hookrightarrow \mathbb{F}_w$ is an injection*

*Proof:* The kernel of the map is the Riemann-Roch space $\mathcal{L}((\alpha - 1)P_\infty - w)$. The degree of the divisor associated with the kernel $deg(\alpha - 1 - deg(w)) = \alpha - 1 - deg(w) < 0$. The dimension of the Riemann-Roch space associated with any divisor of negative degree is zero. Hence the kernel is zero dimensional and hence the map is injective.□

In addition, assume that $\sigma$ is the Frobenius element at $w$. Let $Q$ be the $s'$ variate polynomial that resulted from the interpolation step.

**Lemma 4.2.** *The number of $f \in \mathcal{L}$ that satisfy $Q(f, \sigma(f), \ldots, \sigma^{s'-1}(f)) = 0$ is upper bounded by a polynomial in the block length $N'$.*

*Proof:* Clearly, $Q(f, \sigma(f), \ldots, \sigma^{s'}(f))(w) = Q(f, f^{q^\eta}, \ldots, f^{q^{(s'-1)\eta}})(w)$ as $\sigma$ acts at $w$ as $\sigma(f) \equiv f^{q^\eta} \pmod{w}$, $\forall f \in \mathcal{O}_w$. We define $Q_w := \sum_{i=0}^{s-1} q_i(w) z_1^{a_{i0}} z_2^{a_{i1}} \ldots z_m^{a_{im-1}}$ as the reduction of $Q$ at $w$. If $f \in \mathcal{O}_w$ satisfies $Q(f, \sigma(f), \ldots, \sigma^{s'-1}(f)) = 0$, then $Q(f, f^{q^\eta}, \ldots, f^{q^{(s'-1)\eta}})(w) = 0$. Thus $f(w)$ is a root of $Q_w(z, z^{q^\eta}, \ldots, z^{q^{(s'-1)\eta}})$ over $\mathbb{F}_w$. The degree of $Q(z, z^{q^\eta}, \ldots, z^{q^{(s'-1)\eta}})$ is bounded by $d.q^{(s'-1)\eta}$. Thus the number of roots of $Q_w(z, z^{q^\eta}, \ldots, z^{q^{(s'-1)\eta}})$ in $\mathbb{F}_w$ is bounded by $d.q^{(s'-1)\eta}$. As $\mathcal{L}((\alpha - 1)P_\infty) \hookrightarrow \mathbb{F}_w$ is an injection, the roots $f(w) \in \mathbb{F}_w$ lift to a unique $f \in \mathcal{L}((\alpha - 1)P_\infty)$.□[1]

Thus $d.q^{(s'-1)\eta}$ gives an upper bound on $f \in \mathcal{L}((\alpha-1)P_\infty)$ that satisfy $Q(f, \sigma(f), \ldots, \sigma^{s'-1}(f)) =$

---
[1]Note that for the proof of Lemma 4.2 to be complete, we need to ensure that $Q_w(z, z^{q^\eta}, \ldots, z^{q^{(s'-1)\eta}})$ does not go to zero. Such situations are overcome through a procedure analogous to [9][Lem 6.7] by using the fact that $q^\eta > d$

0. Observe that $d.q^{(s'-1)\eta}$ is polynomial in the block length $N' = \frac{n}{m'}$. This is because, $m$ is a constant fraction of $\frac{N'}{\log(N')}$ and $\alpha \leq n$. Hence the inequality $m\eta \geq \alpha$ holds for an $\eta = C\log_q(n)$ and a large enough constant $C$.

## 4.2 Lifting algorithm to solve the Root Finding Problem:

We describe an algorithm to solve the root finding problem when the order of the automorphism $\sigma$ is small. In this case however the algorithm is much more complicated. We only describe the algorithm for the special case of $m' = m$. The generalization to $m' \neq m$ is straight forward.

We begin by developing some notation about local completions. Let $L_w$ denote the local completion of $L$ at $w$. Let $t$ be a local parameter at $w$. That is $t \in L$ such that $t\mathcal{O}_w = w\mathcal{O}_w$. Every $f \in \mathcal{O}_w$ has an expansion at $w$ of the form $f = \sum_{c=0}^{\infty} f_c t^c \in L_w$. Here $f_c \in \mathcal{O}_w/w\mathcal{O}_w \cong \mathbb{F}_w$. Thus $\mathcal{O}_w$ can be thought of as the ring of infinite power series in $t$, $\mathbb{F}_w[[t]]$. Let $C_{t^c}(f)$ be an alternate notation for the coefficient $f_c$.

The interpolated polynomial $Q(z_1, z_2, \ldots, z_m)$ has degree $d$ and hence can be written as $\sum_\beta a_\beta z_1^{\beta_1} z_2^{\beta_2} \ldots z_m^{\beta_m}$, where $\beta_j \leq d$, $0 \leq j \leq m-1$ and $a_\beta \in L$. Here $\beta$ is used to index the monomials of $Q$. Let $B$ denote the set of all $\beta$. We define $Q_w := \sum_\beta a_\beta(w) z_1^{\beta_1} z_2^{\beta_2} \ldots z_m^{\beta_m}$ as the reduction of $Q$ at $w$.

Elements of $\mathcal{D}_w$ fix $t$ up to a unit. Thus for all $\tau \in \mathcal{D}_w$, $\tau(t) = \zeta t$ where $\zeta$ is a unit in $\mathcal{O}_w$. Clearly being the Frobenius element at $w$, $\sigma$ is contained in $\mathcal{D}_w$, so for all positive $i$ and $j$, $\sigma^i(t^j) = \zeta_{ij} t^j$ where $\zeta_{ij}$ is a unit of $\mathcal{O}_w$. For simplicity of presentation in the discussion below, we will assume that $\sigma(t) = t$. Since $\sigma$ acts on $\gamma \in \mathbb{F}_w$ as $\sigma : \gamma \to \gamma^{q^\eta}$ and fixes $t$, $\sigma$ acts on $\mathcal{O}_w \cong \mathbb{F}_w[[t]]$ as

$$\sigma(\sum_{c=0}^{\infty} f_c t^c) = \sum_{c=0}^{\infty} \sigma(f_c t^c) = \sum_{c=0}^{\infty} f_c^{q^\eta} t^c$$

**Lemma 4.3.** *The linear reduction map* $\phi : \mathcal{L}((\alpha - 1)P_\infty) \hookrightarrow \mathbb{F}_w[[t]]/<t^e>$ *that takes* $f \in \mathcal{L}((\alpha - 1)P_\infty) \subset \mathcal{O}_w$ *to* $\sum_{c=0}^{e} f_c t^c$ *is injective for* $e > \lceil \frac{\alpha}{m\eta} \rceil$.

*Proof:* Let $h \in \mathcal{L}((\alpha - 1)P_\infty)$ be in the kernel of the map. Now $h_c t^c \in \mathcal{L}((\alpha - 1)P_\infty)$. But $t^c$ has $c$ zeros at $w$. Thus $f_c t^c \in \mathcal{L}((\alpha - 1)P_\infty - iw)$. For $i > \lceil \frac{\alpha}{m\eta} \rceil$, $\deg((\alpha - 1)P_\infty - iw) < 0$. Hence $\mathcal{L}((\alpha - 1)P_\infty - iw)$ is zero dimensional and $h_c = 0, c \geq \lceil \frac{\alpha}{m\eta} \rceil$. As $h$ is in the kernel, $h_c = 0, 0 \leq c \leq \lceil \frac{\alpha}{m\eta} \rceil$. Thus $h = 0$. □

We now set $e = \lceil \frac{\alpha}{m\eta} \rceil$. Thus $f \in \mathcal{L}((\alpha - 1)P_\infty)$ can be determined from its truncated expansion $\phi(f) = \sum_{c=0}^{e} f_c t^i$. From the above lemma it is clear that to find the list of messages with sufficient agreement, it suffices to solve the following problem in the local completion.

*Find all* $\phi(f) \in \mathbb{F}_w[[t]]/<t^e>$ *such that* $Q(f, \sigma(f), \ldots, \sigma^{m-1}(f)) = 0$ *in* $\mathbb{F}_w[[t]]$

An algorithm is described in the next section to solve the above problem from which the below result follows. The algorithm depends only on the coefficients $a_{\beta,0}, a_{\beta,1}, \ldots, a_{\beta,e}$. Under the assumption that the received word and the interpolation algorithm induce a distribution where the coefficients $a_{\beta,0}, a_{\beta,1}, \ldots, a_{\beta,e}$ are independent uniformly distributed random variables in $\mathbb{F}_w$, we have the below result.

**Theorem 4.4.** *If* $\{a_{\beta,c}, 0 \leq c \leq e, \beta \in B\}$ *constitute a set of independent, uniformly random elements from* $\mathbb{F}_w$, *then the expected list size is bounded by* $d.q^{(m-1)\eta}$.

A proof of the above theorem is given in the next section.

*Heuristic Assumption:* We assume that for a random received word, the interpolation algorithm maps the received word into $\{a_{\beta,c}, 0 \leq c \leq e, \beta \in B\}$ thereby inducing a distribution wherein $a_{\beta,c}$ are independent, uniformly random elements from $\mathbb{F}_w$.

The heuristic assumption is a natural one because the coefficients of $Q$, $a_\beta \in \mathcal{L}((\alpha - 1)P_\infty)$ are determined as the solution of a linear system that depends on the received word. The linear system is usually close to full rank. This is followed by the reduction of $a_{\beta,c}$ modulo $t^e$.

With this assumption, for a random received word, the expected list size is bounded by $dq^{(m-1)\eta}$, which is a polynomial in the block length.

## 5 Root finding in the Local Completion

We describe an algorithm to determine $\phi(f) \in \mathbb{F}_w[[t]]/<t^e>$ corresponding to $f \in \mathcal{L}((\alpha - 1)P_\infty) \subset \mathbb{F}_w[[t]]$ such that $Q(f, \sigma(f), \ldots, \sigma^{m-1}(f)) = 0$ in $\mathbb{F}_w[[t]]$. As a consequence we have an algorithm that solves the root finding problem for the case of $m$ small compared to the block length. We prove (Theorem 4.4) that the expected number of roots is bounded by a polynomial in the degree of $Q$ and the size of the residue class field $\mathbb{F}_w$ when the coefficients of $Q$ modulo $t^e$ is drawn at random.

We begin by writing down the constraints that $\{f_c\}_{c=0}^e$ corresponding to $\phi(f) = \sum_{c=0}^e f_c t^c$ must satisfy.

**Lemma 5.1.** *For all $f \in \mathbb{F}_w[[t]]$ such that $Q(f, \sigma(f), \ldots, \sigma^{m-1}(f)) = 0$ in $\mathbb{F}_w[[t]]$ and $i \geq 0$,*

$$Q(\sum_{c=0}^{i-1} f_c t^c, \sum_{c=0}^{i-1} f_c^{q^\eta} t^c, \ldots, \sum_{c=0}^{i-1} f_c^{q^{\eta(m-1)}} t^c) \equiv 0 \pmod{t^i}$$

*Proof:* For all $i \geq 0$, we have

$$Q(\sum_{c=0}^\infty f_c t^c, \sigma(\sum_{c=0}^\infty f_c t^c), \ldots, \sigma^{m-1}(\sum_{c=0}^\infty f_c t^c)) \equiv Q(\sum_{c=0}^{i-1} f_c t^c, \sigma(\sum_{c=0}^{i-1} f_c t^c), \ldots, \sigma^{m-1}(\sum_{c=0}^{i-1} f_c t^c)) \pmod{t^i}$$

$$\equiv Q(\sum_{c=0}^{i-1} f_c t^c, \sum_{c=0}^{i-1} f_c^{q^\eta} t^c, \ldots, \sum_{c=0}^{i-1} f_c^{q^{\eta(m-1)}} t^c) \pmod{t^i}$$

$$Q(f, \sigma(f), \sigma^2(f), \ldots, \sigma^{m-1}(f)) = 0 \Rightarrow Q(\sum_{c=0}^\infty f_c t^c, \sigma(\sum_{c=0}^\infty f_c t^c), \ldots, \sigma^{m-1}(\sum_{c=0}^\infty f_c t^c)) \equiv 0 \pmod{t^i}$$

$$\Rightarrow Q(\sum_{c=0}^{i-1} f_c t^c, \sum_{c=0}^{i-1} f_c^{q^\eta} t^c, \ldots, \sum_{c=0}^{i-1} f_c^{q^{\eta(m-1)}} t^c) \equiv 0 \pmod{t^i} \square$$

Further $f \in \mathcal{L}((\alpha - 1)P_\infty)$ is determined by $f \mod t^e$. Hence it suffices to determine $\{f_c\}_{c=0}^e$ such that $Q(\sum_{c=0}^{i-1} f_c t^c, \sum_{c=0}^{i-1} f_c^{q^\eta} t^c, \ldots, \sum_{c=0}^{i-1} f_c^{q^{\eta(m-1)}} t^c) \equiv 0 \pmod{t^e}$. These equations only depend on the coefficients of $Q$ modulo $t^e$.

We begin by determining the list of possible $f_0$. We have $Q(f_0, f_0^{q^\eta}, \ldots, f_0^{q^{\eta(m-1)}}) = 0 \pmod{t}$. Thus $f_0$ is a root of $Q_w(z, z^{q^\eta}, \ldots, z^{q^{\eta(m-1)}})$ in $\mathbb{F}_w$. Hence a list of possible $f_0$ can be enumerated by finding the roots of $Q_w(z, z^{q^\eta}, \ldots, z^{q^{\eta(m-1)}})$ whose degree gives an upper bound of $d.q^{\eta(m-1)}$ on the number of possible $f_0$.

For every fixed $f_0, f_1, \ldots, f_{i-1}$ such that

$$Q(\sum_{c=0}^{i-1} f_c t^c, \sum_{c=0}^{i-1} f_c^{q^\eta} t^c, \ldots, \sum_{c=0}^{i-1} f_c^{q^{\eta(m-1)}} t^c) \equiv 0 \pmod{t^i},$$

we have

$$Q(\sum_{c=0}^{i} f_c t^c, \sum_{c=0}^{i} f_c^{q^\eta} t^c, \ldots, \sum_{c=0}^{i} f_c^{q^{\eta(m-1)}} t^c) \equiv \mu_i t^i \pmod{t^{i+1}}$$

where $\mu_i = C_{t^i}(Q(\sum_{c=0}^{i} f_c t^c, \sum_{c=0}^{i} f_c^{q^\eta} t^c, \ldots, \sum_{c=0}^{i} f_c^{q^{\eta(m-1)}} t^c))$.

Again the set of valid $f_i$ is contained in the set of $f_i$ that satisfy $\mu_i = 0$. Observe that $\mu_i = 0$ is a polynomial equation in $f_0, f_1, \ldots, f_i$. Given that $f_0, f_1, \ldots, f_{i-1}$ are already determined, we can break $\mu_i$ into a polynomial in $f_i$ and a polynomial that does not contain $f_i$. The polynomial in $f_i$ turns out to be very special. It is an additive polynomial whose coefficients depend only on $f_0$ and $a_{\beta,0}$. We now proceed to illustrate this fact and show how this can be exploited to determine $f_i$.

Consider the term

$$(\sum_{c=0}^{\infty} a_{\beta,c} t^c)(\sum_{c=0}^{\infty} f_c t^c)(\sum_{c=0}^{\infty} f_c^{q^\eta} t^c) \ldots (\sum_{c=0}^{\infty} f_c^{q^{(m-1)\eta}} t^c)$$

corresponding to the monomial $a_\beta z_1^{\beta_1} z_2^{\beta_2} \ldots z_m^{\beta_m}$.

The coefficient of $t^i$ that arises from this monomial is

$$a_{\beta,0} \sum_{j=1, \beta_j \neq 0}^{m} f_0^{\lambda_\beta - q^{(j-1)\eta}} f_i^{q^{(j-1)\eta}} + a_{\beta,i} f_0^{\lambda_\beta} + H_{\beta,i}$$

Here $H_{\beta,i}$ depends on $\{a_{\beta,0}, a_{\beta,1}, \ldots, a_{\beta,i-1}, f_0, f_1, \ldots, f_{i-1}\}$ and $\lambda_\beta := \sum_{j=1}^{m} \beta_j q^{(j-1)\eta}$.

By taking the sum over all monomials, we get

$$\mu_i = \sum_\beta a_{\beta,0} \sum_{j=1, \beta_j \neq 0}^{m} f_0^{\lambda_\beta - q^{(j-1)\eta}} f_i^{q^{(j-1)\eta}} + \sum_\beta a_{\beta,i} f_0^{\lambda_\beta} + H_i$$

where $H_i := \sum_\beta H_{\beta,i}$

The term depending on $f_i$ can be rewritten as

$$\sum_\beta a_{\beta,0} \sum_{j=1, \beta_j \neq 0}^{m} f_0^{\lambda_\beta - q^{(j-1)\eta}} f_i^{q^{(j-1)\eta}} = \sum_{j=1}^{m} (\sum_{\beta, \beta_j \neq 0} f_0^{\lambda_\beta - q^{(j-1)\eta}}) f_i^{q^{(j-1)\eta}}$$

Define $F(z) := \sum_{j=1}^{m} (\sum_{\beta, \beta_j \neq 0} f_0^{\lambda_\beta - q^{(j-1)\eta}}) z^{q^{(j-1)\eta}}$. Clearly $F$ is a fixed polynomial independent of $i$ and depends only on $a_{\beta,0}$ and $f_0$.

Now $\mu_i = 0 \Rightarrow F(f_i) + \sum_\beta a_{\beta,i} f_0^{\lambda_\beta} + H_i = 0$. As $f_0, f_1, \ldots, f_{i-1}$ are fixed, we can solve for $f_i$ by finding the roots in $\mathbb{F}_w$ of the polynomial $F(z) + \sum_\beta a_{\beta,i} f_0^{\lambda_\beta} + H_i = 0$.

Observe that the polynomial $F(z) \in \mathbb{F}_w[z]$ is an additive polynomial (or a $q$-polynomial)[6][12] and it is $\mathbb{F}_u$-linear. The roots of $F(z)$ in $\mathbb{F}_w$ thus forms an $\mathbb{F}_u$-linear space. The polynomial $F(z) + \sum_\beta a_{\beta,i} f_0^{\lambda_\beta} + H_i = 0$ is the sum of the additive polynomial $F(z)$ and the constant term $\sum_\beta a_{\beta,i} f_0^{\lambda_\beta} + H_i$. For each $i$, the constant term $\sum_\beta a_{\beta,i} f_0^{\lambda_\beta} + H_i$ is fixed given that $f_0, f_1, \ldots, f_{i-1}$ is fixed. We now state a useful lemma on the structure of the roots polynomial that are the sum of an additive polynomial and a constant.

Let $P(z) \in \mathbb{F}_w[z]$ be an additive polynomial that is $\mathbb{F}_u$−linear. In particular $P$ is of the form $P(z) = \sum_{j=0}^{deg(P)} p_j z^{q^{j\eta}}$, where $p_j \in \mathbb{F}_w$. Let $U$ denote the $\mathbb{F}_u$ linear space of the roots of $P$ in $\mathbb{F}_w$. Let $\delta \in \mathbb{F}_w$ be an arbitrary field element.

**Lemma 5.2.** *If $\gamma_1, \gamma_2 \in \mathbb{F}_w^*$ are two roots of the polynomial $W(z) := P(z) - \delta$, then $\gamma_2 \in \gamma_1 + U$*

*Proof:* The elements $\gamma_1, \gamma_2 \in \mathbb{F}$ are roots of $W$. Thus $P(\gamma_1) = \delta$ and $P(\gamma_2) = \delta \Rightarrow P(\gamma_1) = P(\gamma_2)$. But $P$ is an additive polynomial. Thus $P(\gamma_1) - P(\gamma_2) = 0 \Rightarrow P(\gamma_1 - \gamma_2) = 0 \Rightarrow \gamma_2 \in \gamma_1 + U$.

The converse holds as well. That is, if $\gamma_1$ is a root of $W$, then all the elements of $\gamma_1 + U$ are roots of $W$. Thus the polynomial $W$ either has no roots in $\mathbb{F}_w$ or has exactly $\#U$ roots. Further, $W$ has a root say $\gamma \in \mathbb{F}_w$ if and only if $P(\gamma) = \delta$.

Consider the space of $\mathbb{F}_u$−linear maps from $\mathbb{F}_w$ to $\mathbb{F}_w$. Every such map arises out of the evaluation map of an addiditve polynomial [6]. Let $P(\mathbb{F}_w)$ denote the image of $\mathbb{F}_w$ under the linear map associated with $P$. From the above argument, it is clear that the polynomial $W$ has a root in $\mathbb{F}_w$ if and only if $\delta \in P(\mathbb{F}_w)$.

Define $\delta_i := -\sum_\beta a_{\beta,i} f_0^{\lambda_\beta} - H_i$. The polynomial $F(z) - \delta_i$ has roots in $\mathbb{F}_w$ if and only if $\delta_i \in F(\mathbb{F}_w)$.

This prompts at an iterative procedure that can be used to exhaust the list of all coefficients $\{f_c\}, 0 \leq c \leq e$ that correspond to the messages $f$ in question. We now present the algorithm. Consider a rooted tree with root $r$ and nodes corresponding to elements from $\mathbb{F}_w$.

*The Decoding Algorithm*

- Set of roots of $Q_w(z, z^{q^\eta}, \ldots, z^{q^{\eta(m-1)}})$ in $\mathbb{F}_w$ as the children of the root.

- Compute $U$, the space of roots of $F(z)$ in $\mathbb{F}_w$ and $F(\mathbb{F}_w)$.

- For $i = 1$ to $e$,
  For every path $(r, f_0, f_1, \ldots, f_{i-1})$ do
  
  - If $\delta_i \in F(\mathbb{F}_w)$ with $F(\gamma) = \delta_i$, then set $\gamma + U$ as the children of $f_{i-1}$.

- Lift every $f_0 + f_1 t + \ldots + f_e t^e$ corresponding to a path $(r, f_0, f_1, \ldots, f_e)$ to a function $f \in \mathcal{L}((\alpha - 1)P_\infty)$.

- Output the list of all such functions that have sufficient agreement.

The root finding in the first step can be performed efficiently in time polynomial in the degree of $Q_w(z, z^{q^\eta}, \ldots, z^{q^{\eta(m-1)}})$. The root finding in the second step can be done efficiently by solving a linear system as described in [12][Equation 3.16]. Hence the total running time of the algorithm is bounded by the number of nodes in the tree.

## 5.1 List Size and Running Time of the Algorithm

In this section we present a heuristic argument that shows that for a random received word, the running time of the algorithm as well as the list size grow polynomially in the block length with very high probability.

The list size is clearly upper bounded by the number of leaf nodes at the level $e$ in the tree. The number of choices for $f_0$ is upper bounded by $d.q^{(m-1)\eta}$, which is the degree of the polynomial $Q_w(z, z^{q^\eta}, \ldots, z^{q^{\eta(m-1)}})$. For a fixed $f_0$, we now analyse the number of leaf nodes at level $e$ that are descendents of $f_0$.

Assume that $f_0$ and $a_{\beta,0}$ are fixed. Let $f_{i-1}$ be a descendent of $f_0$ with $f_0, f_1, f_2, \ldots, f_{i-1}$ being the path from $f_0$ to $f_1$. The node $f_{i-1}$ has children if and only if $\delta \in F(\mathbb{F}_w)$. The image $F(\mathbb{F}_w)$ is an $\mathbb{F}_u$ linear space of dimension $m - dim(U)$, where $dim(U)$ is the dimension of $U$. We reiterate that the linear spaces $U$ and $F(\mathbb{F}_w)$ are fixed once $f_0$ is fixed. The probability that a random element in $\mathbb{F}_w$ is in $F(\mathbb{F}_w)$ is $Prob\{\delta_i \in F(\mathbb{F}_w)\} = \frac{\#F(\mathbb{F}_w)}{\#\mathbb{F}_w} = \frac{q^{\eta(m-dim(U))}}{q^{\eta m}} = q^{-\eta dim(U)}$.

The expected number of $f_i$ given $\{f_0, f_1, \ldots, f_{i-1}\}$ is

$$\mathbb{E}(\#f_i | \{f_0, f_1, \ldots, f_{i-1}\}) = \#U.Prob\{\delta_i \in F(\mathbb{F}_w)\} = q^{dim(U)}.q^{-dim(U)} = 1$$

**Lemma 5.3.** *If $\delta_i$ are uniformly random elements from $\mathbb{F}_w$, the expected number of nodes at level $i$ that are descendents of a fixed $f_0$ is bounded by 1*

*Proof:* We prove the above claim by induction. Again, fix $f_0$. The expected number of $f_1$ is thus 1. Assume that the expected number of $f_{i-1}$ that are descendents of $f_0$ is 1 (Induction Hypothesis).

The expected number of $f_i$ is that are descendents of $f_0$ is

$$\sum_{\{f_0,f_1,\ldots,f_{i-1}\}} \mathbb{E}(\#f_i | \{f_0, f_1, \ldots, f_{i-1}\})$$

$$= \sum_{\{f_0,f_1,\ldots,f_{i-1}\}} \#U.Prob\{\delta_i \in F(\mathbb{F}_w)\}$$

$$= \sum_{\{f_0,f_1,\ldots,f_{i-1}\}} 1 = \#\{f_0, f_1, \ldots, f_{i-1}\}$$

But $\#\{f_0, f_1, \ldots, f_{i-1}\}$ is 1 by the induction hypothesis. Thus the expected number of $f_i$ that are descendents of $f_0$ is 1 [2]. $\square$

From the above argument it follows that under the assumption that $\delta_i$ are random elements in $\mathbb{F}_w$, the number of $f_e$ that are descendents of $f_0$ is bounded by 1. Hence the total number of $f_e$ is bounded by the number of $f_0$. Thus the list size is upper bounded by the number of $f_0$. Thus the list size is bounded by $d.q^{(m-1)\eta}$.

From the algorithm description, it is clear that the algorithm depends only on $a_{\beta,0}, a_{\beta,1}, \ldots, a_{\beta,e}$, the coefficients of $Q$ modulo $t^e$. Consider the set of coefficients $\{a_{\beta,c}, 0 \leq c \leq e, \beta \in B\}$. This can be regarded as an element in $\bigoplus_{0 \leq c \leq e, \beta \in B} \mathbb{F}_w$. The interpolation algorithm followed by reduction modulo $t^e$, maps the received word to an element in the finite set $\bigoplus_{0 \leq c \leq e, \beta \in B} \mathbb{F}_w$.

We now present a lemma that relates the distribution of $\{a_{\beta,c}, 0 \leq c \leq e, \beta \in B\}$ to the distribution they induce on $\delta_i$.

**Lemma 5.4.** *For a fixed $f_0$ and $0 < i \leq e$, if $\{a_{\beta,i}, \beta \in B\}$ are independent and uniformly random then $\delta_i$ is a uniformly random variable in $\mathbb{F}_w$.*

---

[2]We have to address the case where $F(z)$ is identically zero. In this case any $f_i \in \mathbb{F}_w$ satisfies $F(f_i) = 0$. However $F(\mathbb{F}_w) = 0$. The probability that $\delta_i = 0$ is $\frac{1}{\#\mathbb{F}_w}$. Thus the expected number of $f_i$ given $f_0, f_1, \ldots, f_{i-1}$ is $\frac{1}{\#\mathbb{F}_w}.\#\mathbb{F}_w = 1$

*Proof:* By definition, $\delta_i = -\sum_\beta a_{\beta,i} f_0^{\lambda_\beta} - H_i$. Consider $H_i$ to be an arbitrary element in $\mathbb{F}_w$. For a fixed $f_0$, $\delta_i$ is a fixed linear combination of $a_{\beta,i}, \beta \in B$ plus an arbitrary constant. Over a finite field a finite linear combination of independent uniformly distributed variables plus an arbitrary element induces the uniform distribution. Thus $\delta_i$ is a uniformly random element in $\mathbb{F}_w$ for every $0 < i \leq e$.□

Consider the case when $\{a_{\beta,c}, 0 \leq c \leq e, \beta \in B\}$ are independent, uniformly random elements from $\mathbb{F}_w$. In this case the constraint that $\{a_{\beta,i}, \beta \in B\}$ are independent uniformly random is clearly satisfied. Thus we have the following theorem.

Finally, Theorem 4.4 follows from lemma 5.3, lemma 5.4 and the fact that the list size is bounded by the number of leaf nodes at level $e$ in the tree.□

## 6 Polynomial List Sizes and A Question on the Existence of Certain Field Extensions

We apply the Folded Algebraic Geometric Code construction (the case of $m' \neq m$) to certain field extensions that have large order automorphisms and solve the root finding problem that arises at the decoder for this special case.

Let $L_a$ be a finite Galois extension of $\mathbb{F}_q(x)$. Assume that we have a sequence of such function fields $L_a$, $a \in \mathbb{Z}^+$ with genus $g(L_a)$ tending to infinity as $a$ grows. The function field sequence $L_a$ is called as asymptotically good if the ratio of the number of rational places in $L_a$ to the genus $g(L_a)$ is bounded away from zero as the genus $g$ grows. This is an informal definition. For a formal definition see [16][V.3.6]. In our context we pose a further restriction and say that $L_a$ is asymptotically good if the ratio of the number of rational places in $L_a$ that resulted out of splitting in the extension (call $n$) to the genus of $L_a$ is bounded away from zero. In addition we require that $L_a$ also have a large order automorphism $\tau \in Gal(L_a/\mathbb{F}_q(x))$.

*Question 6.1:* Does there exist an asymptotically good sequence of function fields $L_a$ such that there exists an element $\tau \in Gal(L_a/\mathbb{F}_q(x))$ whose order $m$ is a constant times $\frac{[L_a:\mathbb{F}_q(x)]}{\log_q([L_a:\mathbb{F}_q(x)])}$ ?

If such an extension exists, the number of rational places in $L_a$ is upper bounded by $q.[L_a : \mathbb{F}_q(x)] = q.\#Gal(L_a/\mathbb{F}_q(x))$. Thus $m$ is a constant fraction of $\frac{N'}{\log_q(N')}$. From section 4.1, we have the following result.

*The codes constructed from $L_a$ are of block length $N'$, rate $R$ over an alphabet of size $q^{m'}$ that can correct $N' - N'(\frac{m'}{m'-s'+1}(R + \frac{m'g}{N'}))^{\frac{s'}{s'+1}}$ errors with a list size bounded by a polynomial in $N'$.*

In addition to being asymptotically good and possesing a large order automorphism, if the tower is asymptotically optimal, then the fraction of errors corrected approaches $1 - R - \epsilon$ for the choice of $m' = \Theta(\frac{1}{\epsilon^2})$ and $s' = \Theta(\frac{1}{\epsilon})$.
A discussion on the existence of asymptotically good towers of function fields with large automorphism follows.

We begin by considering towers where the field at the top is Galois over the rational functional field. The Galois closure of the Garcia-Stichtenoth is one such example. It is interesting to note that the Galois Closure of the Garcia-Stichtenoth towers are optimal as well [18]. Thus in that case the function field on the top of the tower (call $L_a$) is a Galois extension of $\mathbb{F}_{q^2}(x)$. Thus we can hope to use elements of $Gal(L_a/\mathbb{F}_{q^2}(x))$ to fold the code. But the Galois group is non commutative and it is not clear if there exists an element of order comparable to the degree of the extension. In fact, when $q$ is prime, the Galois group is $\bigoplus_{i=0}^{a} \mathbb{Z}/q\mathbb{Z}$. In this case no such large order automorphisms exist and all elements have order at most $q$.

There certainly exists geometric extensions with large automorphisms. For instance, there exists cyclic extensions (Galois Group is cyclic) over $\mathbb{F}_q(x)$ of arbitrarily large degree, when the degree of the extension is a power of $q$. These are called as cyclotomic function fields [15][chap 12],[6][chap 3] and are generated by adjoining to $\mathbb{F}_q(x)$, a torsion submodule of the division points of a Carlitz module. However, such extensions do not posses enough places of small degree as illustrated below. The prospect of using cyclotomic function fields in folded codes was inspired by a communication with Venkatesan Guruswami[7] for which we thank him.

## 6.1 Cyclotomic Function Fields

Cyclotomic function fields are certain geometric extensions of function fields where the Galois group of the extension is cyclic. A description of cyclotomic function fields follows.The notation and definitions are based on [15][chap 12] .

Let $k$ be a function field of characteristic $p$ with $\mathbb{F}_q$ as the field of constants. Let $\tau$ denote the $q$−th power map. Let $k<\tau>$ denote the ring of twisted polynomials over $k$ with the commutation rule $\tau h = h^q \tau$, $\forall h \in k$. This is the ring of additive endomorphisms of $\bar{k}$ (the algebraic closure of $k$) that fix $\mathbb{F}_q$.

We now consider a special case. Set $A = \mathbb{F}_q[T]$ and $k = \mathbb{F}_q(T)$.

A Drinfeld module for $A$ is an $\mathbb{F}_q$ algebra homomorphism $\rho : A \to k<\tau>$ such that $\forall a \in A$, the constant term of the image $\rho_a$ is $a$. Further, to ensure non-triviality, for at least one $a \in A$, the image $\rho_a \notin k$. The homomorphism $\rho$ gives $\bar{k}$ an $A$−modules structure by defining the multiplication $a.u = \rho_a(u), \forall a \in A, \forall u \in \bar{k}$. Consider the module $\Lambda_\rho[a] := \{\lambda \in \bar{k} | \rho_a(\lambda) = 0\}$. For every non zero $a \in A$, $\Lambda_\rho[a] \cong A/aA \oplus A/aA \oplus \ldots \oplus A/aA$ ($r$ times). Here $r$ is the rank of the Drinfeld module.[15][12.4]

Let $k_{\rho,a} := k(\Lambda_\rho[a])$ denote the extension obtained by adjoining the elements of $\Lambda_\rho[a]$ to $k$. Such extensions are Galois extensions [15].

A rank-one Drinfeld module with $\rho_T = T + \tau$ is called as a Carlitz module. From now on, we confine our attention to Carlitz modules. In the case of Carlitz modules, $Gal(k_{\rho,a}/k)$ is abelian [15][12.5]. This abelian extension $k_{\rho,a}/k$ is called as a cyclotomic function field. In this case the degree of the extension is $\Phi(a)$, the number of non zero polynomials of degree less than the degree of $a$ and relatively prime to $a$. Let $deg(a)$ denote the degree of $a$ as a polynomial in $A$. In fact $Gal(k_{\rho,a}/k)$ is cyclic if $a$ is irreducible.

The following theorem describes the splitting behavior of places in the extension $k_{\rho,a}/k$. Let $u$ be a place in $k$ that is not a place at infinity. Let $f$ be the smallest integer such that $u^f = 1 \mod a$.

**Theorem 6.1.** *The place $u$ factors into $\Phi(a)/f$ places in $k_{\rho,a}$. [15][12.10].*

**Lemma 6.2.** *Every place in $k_{\rho,a}$ has degree at least $deg(a)$, except possibly for places at infinity.*

*Proof:* The degree of a place $v \in k_{\rho,a}$ lying above $u \in k$ is $f.deg(u)$. But $u^f = 1 \mod a \implies deg(u).f \geq deg(a) \implies deg(v) \geq deg(a)$.$\square$

Thus every place in $k_{\rho,a}$ (apart from places at infinity) has degree at least $deg(a)$. Hence any evaluation based code on the function field $k_{\rho,a}$ that uses places away from infinity for evaluation has an alphabet size of at least $q^{deg(a)}$. This is because the size of the residue class fields at these places is at least $q^{deg(a)}$.

The degree of extension $[k_{\rho,a} : k]$ is upper bounded by $q^{deg(a)}$. Thus the order of any element in $Gal(k_{\rho,a}/k)$ is upper bounded by $q^{deg(a)}$. Consider the case where the cyclotomic function field $k_{\rho,a}$ is used in a folded construction. Suppose places of degree $d (> deg(a))$ are used for evaluation. The block length in this case is at most the number of places of degree $d$. The number of places of degree at most $d$ in $k$ is bounded by $q^d$. At best all these places split, giving $q^d[k_{\rho,a}, k]$ places of degree at most $d$ in $k_{\rho,a}$.

Thus $q^d[k_{\rho,a}, k] \leq q^d q^{deg(a)}$ is an upper bound on the block length. The alphabet size is at least $q^{dm'}$, where $m' \geq 2$ is the folding parameter. Thus the alphabet size of these codes is at least $q^{2d}$. However, the block length is bounded by $q^d q^{deg(a)} \leq q^{2d}$. Thus the alphabet size is at least as big as the block length[3].

One major motivation for generalizing Folded Reed-Solomon codes to Folded Algebraic-Geometric codes is to find codes over an alphabet independent of the block length. But folded codes defined on cyclotomic function fields do not improve on Folded Reed Solomon codes in terms of alphabet size. Recently, Guruswami [8] overcame this obstacle by considering certain special subfields of the cyclomic fields thereby achieving an alphabet size that is logarithmic in the block length.

# 7 Folded Codes from Garcia-Stichtenoth Towers

Garcia and Stichtenoth described [4] function field towers that are asymptotically optimal. That is they attain the Drinfeld-Vladut bound. We apply the construction with $m' \neq m$ to these towers of function fields. We state the below theorems quantifying the error correction performance of these codes.

**Theorem 7.1.** *The folded codes from Garcia-Stichtenoth towers of rate $R$, block length $N$ over an alphabet of size $q^{2m}$ can correct $N(1 - (R + \frac{m}{q-1})^{\frac{m}{m+1}})$ errors.*

The expected list size bounded by $N^{O(m)}$ under the heuristic assumption

**Theorem 7.2.** *The Folded codes from Garcia-Stichtenoth towers of rate $R$ can correct up to a fraction of $1 - R - \epsilon$ errors over an alphabet of size $(\frac{1}{\epsilon^2})^{O(\frac{1}{\epsilon})}$ independent of the size of the block length.*

The expected list size is bounded by $N^{O(\frac{1}{\epsilon})}$ under the heuristic assumption.

These are towers defined as a sequence of Artin-Schreier extensions. The base field is the finite field $\mathbb{F}_{q^2}$, where $q$ is a prime power. $F_0$ is the rational function field $F_0 = \mathbb{F}_{q^2}(x)$.

$$F_i = F_{i-1}(x_n)$$
$$x_i^q + x_i = \frac{x_{i-1}^q}{x_{i-1}^{i-1} + 1}, 1 \leq i \leq a.$$

The splitting behavior of places in the tower is critical to our code construction and is completely described in [1]. Let $S_{sp}$ denote the set of all places in $F_a$, that resulted out of complete splitting in the extension $F_a/F_{a-1}$. Let $P_\theta^0, \theta \in F_0$ denote the unique place in $F_0$ that is the zero of $x_0 - \theta$. Let $\Omega := \{\beta \in F_0 : \beta^q + \beta = 0$ denote the set of $q$ trace zero elements in $F_0$. The places $P_\theta^0, \theta \in F_0 \setminus \Omega$ completely split in the extension $F_a/F_0$. So the number of places in $S_{sp}$ is at least $q^b(q^2 - q)$. The extension $F_i/F_{i-1}$ is Galois, but unfortunately the extension $F_a/F_0$ is not a Galois extension. So we use automorphisms in the Galois Group of the extension $F_a/F_{a-1}$ to fold the codes. The Galois Group $Gal(F_i/F_{i-1})$ is isomorphic to $\Omega$, the additive group of all trace zero elements in $\mathbb{F}_{q^2}$ with trace taken down to $\mathbb{F}_q$. In particular, any non trivial element $\sigma \in Gal(F_a/F_{a-1})$ has order $m$ that equals the characteristic $p$ of the finite field $\mathbb{F}_q$. The genus $g$ of the function field $F_a$ is $(q^{\frac{a}{2}}-1)(q^{\frac{a+2}{2}}-1)$ if $a$ is even and $(q^{\frac{a+1}{2}}-1)^2$ if $a$ is odd. In either case the genus is approximately $q^{a+1}$. The point at infinity in $F_0$ is completely ramified in throughout the tower and there is a unique place at infinity $P_\infty \in F_a$ of degree 1. As $P_\infty$ is totally ramified, $P_\infty$ is fixed by any element of $Gal(F_a/F_{a-1})$.

---
[3]The sum of the degrees of places at infinity in $k_{\rho,a}$ is at most $q^{deg(a)}$. The genus of $k_{\rho,a}$ is larger than $q^{deg}(a)$. Thus if only places at infinity are used as evaluation places, the message space of the code is trivial.

The automorphism $\sigma$ is used to fold the places $= S_{sp}$. By evaluating $\mathcal{L}((\alpha - 1)P_\infty)$ at $P_s$, we get a folded algebraic geometric code with $n = q^b(q^2 - q)$ and a folding parameter of $m$. Observe that by increasing $a$ we can make $n$ arbitrarily large compared to $m$.

Thus the block length of the resulting code is $N = \frac{q^a(q^2-q)}{m}$. The dimension of the code $k = dim(\mathcal{L}((\alpha-1)P_\infty))$. If $\alpha - 1 \geq 2g - 2$, then $k = \alpha - g$. The code is over an alphabet of size $q^{2m}$ and under our heuristic can be decoded if the agreement $T$ is at least $\sqrt[m+1]{N(\alpha-1)^m}$ with expected list size bounded by $d.q^{2(m-1)\eta}$. Thus the number of errors that can be corrected is $N - \sqrt[m+1]{N(\alpha-1)^m} = N(1 - (\frac{k+g}{N})^{\frac{m}{m+1}}) = N(1 - (\frac{k}{N} + \frac{m}{q-1})^{\frac{m}{m+1}})$

Observe that $n/g$ tends to $q - 1$ as $g$ grows. Here $m$ equals $p$, the characteristic of the finite field $\mathbb{F}_{q^2}$. Theorem 7.1 follows

The expected list size bounded by $N^{O(m)}$ under the heuristic assumption

Observe that the Folded codes from Garcia-Stichtenoth towers of rate $R$ can correct up to a fraction of $1 - R - \epsilon$ errors when $m = p = O(\frac{1}{\epsilon})$ and $q = p^b$ with $b > 2$. This is the optimum tradeoff in terms of rate and error correction[3]. Thus if $q = p^2$, we can achieve the optimum rate-error correction tradeoff over an alphabet of size $(\frac{1}{\epsilon^2})^{O(\frac{1}{\epsilon})}$ independent of the size of the block length. Theorem 7.2 follows.

# References

[1] Aleshnikov, I., Kumar, P.V., Shum, K.W., Stichtenoth, H. "On the splitting of places in a tower of function fields meetingthe Drinfeld-Vladut bound", *IEEE Transactions on Information Theory, Volume 47, Issue 4, May 2001 Page(s):1613 - 1619*

[2] Drinfeld, V. G., Vladut, S. G. "Number of Points of an Algebraic Curve". *Func. Anal.17, 5354 (1983)*

[3] Peter Elias, "Error-correcting codes for list decoding", *IEEE Transactions on Information Theory*, vol. 37, pp. 512, 1991.

[4] Arnoldo Garcia, Henning Stichtenoth, "Algebraic function fields over finite fields with many rational places". *IEEE Transactions on Information Theory 41(6): 1548-1563 (1995).*

[5] V D Goppa "Geometry and Codes" *in Mathematics and its Applications. Kluwer Academic Press*

[6] David Goss, "Basic Structures of Function Field Arithmetic". *A Series of modern Surveys in Mathematics, Vol 35. Springer-Verlag. 1996*

[7] Venkatesan Guruswami, "Personal Communication". 2008

[8] Venkatesan Guruswami, "Artin automorphisms, Cyclotomic function fields, and Folded list-decodable codes", http://arxiv.org/abs/0811.4139

[9] Venkatesan Guruswami, Anindya C. Patthak, "Correlated Algebraic-Geometric Codes: Improved List Decoding over Bounded Alphabets". *Mathematics of Computation, 77(2008), 447-473.*

[10] Venkatesan Guruswami, Atri Rudra, "Explicit codes achieving list decoding capacity: Error-correction up to the Singleton bound", *IEEE Trans. on Info. Theory, 54(1), Jan 2008.*

[11] Venkatesan Guruswami, Madhu Sudan, "Improved decoding of Reed-Solomon and algebraic-geometric codes " *IEEE Trans. on Info. Theory*

[12] Rudolf Lidl ,Harald Niederreiter, "Finite Fields",*Vol. 20 in the Encyclopedia of Mathematics and its Applications, Addison-Wesley.*

[13] V Kumar Murty, J Scherk, "Effective versions of the Chebotarev density theorem for function fields",*C. R. Acad. Sci. (Paris), 319 (1994)*

[14] Farzard Parvaresh, Alexander Vardy, "Correcting errors beyond the Guruswami-Sudan radius in polynomial time" *IEEE Foundation of Computer Science 2005*

[15] Michael Rosen, "Number Theory in Function Fields". *Graduate Texts in Mathematics*

[16] Henning Stichtenoth, "Algebraic Function Fields and Codes". *Series: Universitext, Springer-Verlag . 1993, X, 260 p.*

[17] Tsfasman, M. A., Vladut, S. G., Zink, T. " Modular Curves, Shimura Curves and Goppa Codes, better than the Varshamov-Gilbert Bound". *Math. Nachr.109, 2128 (1982)*

[18] Alexey Zaytsev, "The Galois closure of the Garcia-Stichtenoth tower", arXiv:math/0504431v1 [math.AG].

[19] ] V.V. Zyablov, M.S. Pinsker, " List cascade decoding", *Probl. Pered. Inform. 17 (1981) 29-33.*